# Optimal Operation of Reconfigurable Active Distribution Networks Aiming at Resiliency Improvement

Saeed Behzadi
*Department of Electrical Engineering, Faculty of Engineeing, University of Zanjan, Zanjan, Iran*
saeedbehzadi@znu.ac.ir

Amir Bagheri
*Department of Electrical Engineering, Faculty of Engineeing, University of Zanjan, Zanjan, Iran*
a.bagheri@znu.ac.ir

Abbas Rabiee
*Department of Electrical Engineering, Faculty of Engineeing, University of Zanjan, Zanjan, Iran*
Rabiee@znu.ac.ir

***Abstract*— As natural disasters bring about power outage and financial losses, network resiliency is an important challenge for distribution network operators (DNOs). On the other side, power loss reduction during normal operating condition is a major concern of DNOs. In this paper, optimal scheduling of active distribution network (ADN) is addressed through simultaneous minimization of power loss in normal condition and load shedding in critical condition after natural disasters. A new formulation is developed for the network reconfiguration to optimize the system operation in both normal and emergency conditions in the presence of conventional and renewable-energy-based distributed generation (DG) as well as energy storage systems (ESSs). The line flow based (LFB) algorithm is used for the AC power flow calculations, and all the developed relations have been convexified to construct a mixed-integer quadratically-constrained programming (MIQCP) optimization model. The simulations have been implemented on the IEEE 33-bus system in GAMS, and the results are investigated.**

## I. NOMENCLATURE

| | |
|---|---|
| $i\ /\ j$ | Buses |
| $T$ | Time |
| $C$ | System operating mode |
| $\ell$ | Line between buses *i* and *j* |
| $\Omega_N$ | Set of buses |
| $\Omega_L$ | Set of branches |
| $\Omega_T$ | Set of time |
| $\Omega_C$ | Set of system operating mode |
| $\Omega_S$ | Set of slack buses |
| $\Omega_{DG}$ | Set of candidate buses for installing a synchronous generator |
| $\Omega_{ESS}$ | Set of candidate buses for installing energy storage systems |
| $\Omega_{WT}$ | Set of candidate buses for installing wind turbines |
| $R_\ell^{Line}$ | Line resistance |
| $X_\ell^{Line}$ | Line reactance |
| $P_i^L\ /\ Q_i^L$ | Peak value of active/reactive load demand |
| $\lambda_t$ | Daily load factor |
| $PF_c$ | Generation power factor of synchronous DGs |
| $bigM$ | A big number |
| $V_{Min}\ /\ V_{Max}$ | Minimum/Maximum voltage of buses |
| $I_\ell^{Max}$ | Maximum capacity of lines (in Ampere) |
| $P_{Max}^G\ /\ Q_{Max}^G$ | Maximum active/reactive power injected into the system from the upstream network |
| $P_{Min}^{net}\ /\ P_{Max}^{net}$ | Minimum/maximum active power passing through lines |
| $Q_{Min}^{net}\ /\ Q_{Max}^{net}$ | Minimum/maximum reactive power passing through lines |
| $H_{Min}\ /\ H_{Max}$ | Minimum/maximum hypothetical active power that can be passed through lines |
| $G_{Max}$ | Maximum hypothetical active power |
| $S_{Max}^{DG}$ | Capacity of synchronous generators |
| $SOC^{Max}$ | Maximum energy storage capacity of ESS |
| $P_{Max}^{ESS}$ | Maximum active power of ESS |
| $EFC\ /\ EFD$ | Charge/discharge efficiency of ESS |
| $V_t^{wind}$ | Wind speed |
| $P_{rated}$ | Rated power of wind turbines |
| $P_{i,c,t}^{WTa}$ | Maximum active power capacity of wind generators |
| $V_{c_{in}}\ /\ V_{c_{out}}$ | Cut-in/cut-out speeds of wind turbine |
| $V_{rated}$ | Rated speed of wind turbine |
| $OF$ | Main objective function |
| $OF_1,\ OF_2$ | Objective function for operating modes of $c_1$ and $c_2$ |
| $OF_1^{optimum}$ | Optimal value of the first objective function without the influence of second objective function |
| $OF_2^{optimum}$ | Optimal value of the second objective function without the influence of first objective function |
| $RI$ | System resilience index |
| $V_{i,c,t}$ | Bus voltage magnitude |
| $I_{\ell,c,t}$ | Line current |
| $U_{i,c,t}$ | Square of bus voltage |
| $J_{\ell,c,t}$ | Square of line current |
| $P_{i,c,t}^{Lsh}\ /\ Q_{i,c,t}^{Lsh}$ | Active/reactive power of load shedding |
| $P_{i,c,t}^{Lsh_a}$ | Hypothetical active power of load shedding |
| $P^{L_{total}}$ | Total system load |
| $P_{i,j,c,t}^{net}\ /\ Q_{i,j,c,t}^{net}$ | Active/reactive power passing through the lines |
| $P_{i,c,t}^G\ /\ Q_{i,c,t}^G$ | Active/reactive power injected into the system from the upstream network |
| $P_{i,c,t}^{DG}\ /\ Q_{i,c,t}^{DG}$ | Active/reactive power injected into the |
| | system by synchronous generators |
| $P^{WT}_{i,c,t}$ | Active power injected into the system by wind generators |
| $X_{i,j,c}$ | Binary variable of connection/disconnection of lines |
| $r^{pos}_{i,c,t}$ / $r^{neg}_{i,c,t}$ | Positive/negative binary variable used for linearization of absolute function |
| $ich_{i,t}$ / $idch_{i,t}$ | Charge/discharge binary variable of ESS |
| $G_{i,c,t}$ | Hypothetical injected active power |
| $H_{i,j,c,t}$ | Hypothetical active power passing through lines |
| $DG^{a}_{i,c}$ | Hypothetical binary variable for DG unit |
| $SOC_{i,c,t}$ | State of charge for ESS unit |
| $PC_{i,c,t}$ / $PD_{i,c,t}$ | Active charging/discharging capacity of ESS |

## II. INTRODUCTION

Nowadays, due to climate change, occurrence of severe natural disasters including storms, floods, droughts, and etc. have been intensified in many countries [1]. These extreme weather events usually have low probability of occurrence, while they can severely damage power system infrastructures resulting in long-term power outages in distribution networks [2]. In this regard, "*power system resilience*" is a measure for evaluating the ability of a system to withstand against significant power outages caused by natural disasters or intentional attacks, and then restoration of loads after occurring these events [3]. Prevention, survivability, and recovery are three main parts of the power system resiliency [4]. The traditional way for resiliency improvement is the power system infrastructure reinforcement including construction of additional lines/substation [5], enhancing towers [6], etc. However, these approaches are high-cost solutions requiring long construction time. In addition, they are subject to right-of-way and environmental concerns. In today's distribution systems, integration of distributed energy resources (DERs) including conventional and renewable-energy-based distributed generation (DG) as well as ESS units have changed the conventional distribution networks to ADNs [7, 8]. DER units help to improve operation of ADNs in both normal and emergency operating conditions [9]. Optimal scheduling of these units can significantly reduce network power loss, improve voltage profile, and release network equipment capacity [10, 11]. In addition, under emergency conditions, when some parts of the network are affected by natural disasters, DERs can act as backup energy sources to minimize loads interruption and maximize the resiliency index [12]. Another alternative that the DNOs can utilize for optimizing network operation in normal condition and enhancing the system resilience in critical situations is the optimal network reconfiguration and optimal micro-grid formation [13, 14]. By optimal opening/closing plan of tie lines and switches, the DNO will optimally operate the distribution system in either normal or emergency conditions. Several studies have addressed optimal scheduling of distribution networks using DER and reconfiguration. Some of these researches address optimal network operation in normal conditions, and some other dedicate to improvement of system resiliency under emergency conditions after natural disasters. In order to restore critical loads and improve network resilience after natural disasters, a chance-constrained stochastic method is proposed in [15] in which the micro-grids formation strategy is utilized. In addition, the effects of fault location, available generation resources, and load priority on the restoration approach are investigated. In [16], to increase resilience of system and decrease the value of load shedding, a two-stage stochastic optimization model is proposed for wind turbine allocation and network reconfiguration after events occurrence. Also, the effects of uncertainty are taken into account. The proposed model is mixed integer non-linear programming (MINLP) and has non-convex nature. It should be noted that the solution provided for the non-convex model is usually non-optimal. Also, the ESS units are not present in the formulations of [16]. Ref. [17] has improved resilience of distribution network in dealing with severe natural disasters by utilizing DG and ESS as well as micro-grid formation. The proposed model is a mixed-integer linear programming (MILP) one in which the linearized power flow equations are employed. Ref. [18] proposes a two-stage stochastic MILP model in which the network can be divided into dynamic micro-grids and benefit from internal combustion engines (ICE) cars to improve resilience index. The results prove that the amount of restored load is directly affected by the buses corresponding to ICE cars. The power loss reduction in normal operating condition has not been addressed in [17] and [18]. A model based on combination of reconfiguration and micro-grid formation for improving critical load restoration capability in distribution system after natural disasters is presented in [19]. Also, the proposed objective function considers minimization of total system loss and number of switching actions. The solution approach is based on a heuristic method using spanning tree search algorithm. This study is in the absence of ESS units. Also, the heuristic approaches cannot guarantee the optimality. A bi-level optimization problem is developed in [20] with two conflicting objectives including maximizing network resilience and minimizing number of switching operations. The loads priority as an important characteristic of modern distribution systems is taken into consideration. The model is optimized using a bi-level genetic algorithm in an MINLP model.

The literature review shows that there is no work around network resiliency improvement along with optimizing normal operating conditions using a convexified model. In this paper, optimal scheduling of ADN is fulfilled through simultaneous minimization of power loss in normal condition and load shedding in critical condition after natural disasters. A new convex formulation is developed which employs reconfiguration and optimal network configuration in order to optimize the system operation in both normal and emergency conditions in the presence of DER units. The AC power calculations using LFB technique is used to develop a MIQCP optimization model which will be solved in GAMS.

## III. PROBLEM FORMULATIONS

### A. Objective Function

A major concern of DNOs is reducing active power losses in distribution networks. On the other hand, in critical conditions after natural disasters, minimization of load shedding is considered as the main goal. In this paper, the proposed objective function includes two terms as (1). The first term is associated with the power loss reduction according to (2), and the second one is related to total load shedding in emergency mode expressed as (3). The normal and emergency modes of the network are denoted by $c_1$ and $c_2$, respectively. For convenience, the parameters of the relations have been defined in the nomenclature.

$$OF = Min\left\{\frac{OF_1}{OF_1^{optimum}} + \frac{OF_2}{OF_2^{optimum}}\right\} \tag{1}$$

$$OF_1 = \sum_{t\in\Omega_T}\sum_{\ell\in\Omega_L} R_\ell^{Line} J_{\ell,c,t} \quad ;\forall c\in c_1 \tag{2}$$

$$OF_2 = \sum_{t\in\Omega_T}\sum_{i\in\Omega_N} P_{i,c,t}^{Lsh} \quad ;\forall c\in c_2 \tag{3}$$

### B. Problem constraints

#### 1) Power flow equations

In this paper, a convexified model of AC power flow proposed as (4)-(19). This formulation is called as line-flow-based (LFB) technique for the power flow.

$$\forall i,j\in\Omega_N, \forall t\in\Omega_T, \forall \ell\in\Omega_L, \forall c\in\Omega_C:$$

$$\begin{aligned}&\sum_{j\in\Omega_N} A_{\ell i} P_{i,j,c,t}^{net} = P_{i,c,t}^{G} + P_{i,c,t}^{DG} + P_{i,c,t}^{WT} + P_{i,c,t}^{Lsh}\\ &+PD_{i,c,t} - PC_{i,c,t} - \lambda_t P_i^L - \sum_{j\in\Omega_N} B_{\ell i} R_\ell^{Line} J_{\ell,c,t}\end{aligned} \tag{4}$$

$$\begin{aligned}&\sum_{j\in\Omega_N} A_{\ell i} Q_{i,j,c,t}^{net} = Q_{i,c,t}^{G} + Q_{i,c,t}^{DG} + Q_{i,c,t}^{Lsh} - \lambda_t Q_i^L\\ &-\sum_{j\in\Omega_N} B_{\ell i} X_\ell^{Line} J_{\ell,c,t}\end{aligned} \tag{5}$$

$$\begin{aligned}&-\left(1-X_{i,j,c}\right)bigM \le U_{j,c,t} + 2\left(R_\ell^{Line} P_{i,j,c,t}^{net} + X_\ell^{Line} Q_{i,j,c,t}^{net}\right)\\ &-U_{i,c,t} + \left[\left(R_\ell^{Line}\right)^2 + \left(X_\ell^{Line}\right)^2\right] J_{\ell,c,t} \le \left(1-X_{i,j,c}\right)bigM\end{aligned} \tag{6}$$

$$\left(P_{i,j,c,t}^{net}\right)^2 + \left(Q_{i,j,c,t}^{net}\right)^2 \le J_{\ell,c,t} U_{j,c,t} \tag{7}$$

$$P_{i,j,c,t}^{net} = -P_{j,i,c,t}^{net} \tag{8}$$

$$Q_{i,j,c,t}^{net} = -Q_{j,i,c,t}^{net} \tag{9}$$

$$-X_{i,j,c} P_{Min}^{net} \le P_{i,j,c,t}^{net} \le X_{i,j,c} P_{Max}^{net} \tag{10}$$

$$-X_{i,j,c} Q_{Min}^{net} \le Q_{i,j,c,t}^{net} \le X_{i,j,c} Q_{Max}^{net} \tag{11}$$

$$X_{i,j,c} = X_{j,i,c} \tag{12}$$

$$\begin{cases} V_{i,c,t} = 1 & ;\forall c\in c_1, i\in\Omega_S\\ V_{Min} \le V_{i,c,t} \le V_{Max} & \end{cases} \tag{13}$$

$$U_{i,c,t} = \left(V_{i,c,t}\right)^2 \tag{14}$$

$$\left(V_{Min}\right)^2 \le U_{i,c,t} \le \left(V_{Max}\right)^2 \tag{15}$$

$$0 \le J_{\ell,c,t} \le X_{i,j,c}\left(I_\ell^{Max}\right)^2 \tag{16}$$

$$J_{\ell,c,t} = \left(I_{\ell,c,t}\right)^2 \tag{17}$$

$$\begin{cases} 0 \le P_{i,c,t}^{G} \le P_{Max}^{G} & ;\forall i\in\Omega_S,\ c\in c_1\\ P_{i,c,t}^{G} = 0 & ;\quad otherwise \end{cases} \tag{18}$$

$$\begin{cases} 0 \le Q_{i,c,t}^{G} \le Q_{Max}^{G} & ;\forall i\in\Omega_S,\ c\in c_1\\ Q_{i,c,t}^{G} = 0 & ;\quad otherwise \end{cases} \tag{19}$$

In (4) and (5) which represent active and reactive power balance at the buses ($i$) in different time periods ($t$) and operating conditions ($c$), $A_{\ell i}$ is the bus-line matrix which equals to 1 if the sending bus of line $\ell$ is bus $i$, and equals to -1, if the receiving bus of line $\ell$ is bus $i$. Otherwise, $A_{\ell i}$ will be zero. On the other hand, in $A_{\ell i}$ if 0 is placed instead of 1, $B_{\ell i}$ is formed. Relation (6) represents the voltage drop along a branch considering the network reconfiguration binary variable ($X_{i,j,c}$), and (7) shows the convexified relation between current, voltage and powers of each branch. Relations (10), (11), and (16) guarantees when each line is disconnected ($X_{i,j,c}=0$) no current nor active and reactive powers should pass through the line, and when line is connected ($X_{i,j,c}=1$), its current limits must be satisfied.

#### 2) Radiality constraint

According to (20)-(28), to form the distribution system with a radial structure, a hypothetical power flow is considered in this paper. The related parameters have been given in the nomenclature.

$$\forall i,j\in\Omega_N, \forall t\in\Omega_T, \forall \ell\in\Omega_L, c\in\Omega_C:$$

$$\sum_{\ell\in\Omega_L} X_{i,j,c} = 2\left[\left(\sum_{i\in\Omega_N} i\right)-1\right] \tag{20}$$

$$G_{i,c,t} + P_{i,c,t}^{Lsh_a} - \lambda_t P_i^L = \sum_{j\in\Omega_N} H_{i,j,c,t} \tag{21}$$

$$H_{i,j,c,t} + H_{j,i,c,t} = 0 \tag{22}$$

$$X_{i,j,c} H_{Min} \le H_{i,j,c,t} \le X_{i,j,c} H_{Max} \tag{23}$$

$$\begin{cases} 0 \le G_{i,c,t} \le G_{Max} & ;\forall i\in\Omega_S, \forall c\in c_1\\ 0 \le G_{i,c,t} \le G_{Max} DG_{i,c}^{a} & ;\forall i\in\Omega_{DG}, \forall c\in c_2\\ G_{i,c,t} = 0 & ;\quad otherwise \end{cases} \tag{24}$$

$$0 \le P_{i,c,t}^{Lsh_a} \le 0.99\lambda_t P_i^L \tag{25}$$

$$DG_{i,c}^{a} = 0 \quad ;\forall c\in c_1 \tag{26}$$

$$\sum_{i\in\Omega_{DG}} DG_{i,c}^{a} = 1 \quad ;\forall c\in c_2 \tag{27}$$

#### 3) Load shedding constraints

In order to determine amount of load shedding which is occurred only in the emergency condition, the following constraints must be satisfied.

$$\forall i\in\Omega_N, \forall t\in\Omega_T, \forall c\in c_2:$$

$$Q_{i,c,t}^{Lsh} = \left(\frac{\lambda_t Q_i^L}{\lambda_t P_i^L}\right) P_{i,c,t}^{Lsh} \tag{28}$$

$$0 \le P_{i,c,t}^{Lsh} \le \lambda_t P_i^L \tag{29}$$

$$0 \le Q_{i,c,t}^{Lsh} \le \lambda_t Q_i^L \tag{30}$$

$$\begin{cases} P_{i,c,t}^{Lsh} = 0 & \forall c\in c_1\\ Q_{i,c,t}^{Lsh} = 0 & \forall c\in c_1 \end{cases} \tag{31}$$

#### 4) Constraints of synchronous DGs

The relation between active and reactive powers as well as generation limit are given as (32) and (33).

$$\forall i\in\Omega_{DG}, \forall t\in\Omega_T, \forall c\in\Omega_C:$$

$$-tan\left(cos^{-1}\left(PF_c\right)\right)P_{i,c,t}^{DG} \le Q_{i,c,t}^{DG} \le tan\left(cos^{-1}\left(PF_c\right)\right)P_{i,c,t}^{DG} \tag{32}$$

$$\left(P_{i,c,t}^{DG}\right)^2+\left(Q_{i,c,t}^{DG}\right)^2 \le \left(S_{Max}^{DG}\right)^2 \tag{33}$$

5) *ESS and wind turbine units*

The relations regarding charge/discharge states of ESS units are given in (34)-(39). Also, (40) and (41) exhibit the power-speed relation of wind turbine and its generation limit, respectively.

$\forall i \in \Omega_{ESS}, \forall t \in \Omega_T, \forall c \in \Omega_C$ :

$$SOC_{i,c,t} = SOC_{i,c,(t-1)} + \left(PC_{i,c,t} EFC\right) - \left(\frac{PD_{i,c,t}}{EFD}\right) \tag{34}$$

$$0 < SOC_{i,c,t} < SOC^{Max} \tag{35}$$

$$0 \le PC_{i,c,t} \le ich_{i,t} P_{Max}^{ESS} \tag{36}$$

$$0 \le PD_{i,c,t} \le idch_{i,t} P_{Max}^{ESS} \tag{37}$$

$$ich_{i,t} + idch_{i,t} \le 1 \tag{38}$$

$$SOC_{i,c,t_{24}} = SOC_{i,c,t_0} \tag{39}$$

$\forall i \in \Omega_{WT}, \forall t \in \Omega_T, \forall c \in \Omega_C$ :

$$P_{i,c,t}^{WTa} = \begin{cases} 0 & ;\forall V_t^{wind} < V_{c_{in}} \\ P_{rated}\left(\dfrac{V_t^{wind} - V_{c_{in}}}{V_{rated} - V_{c_{in}}}\right) & ;\forall V_{c_{in}} \le V_t^{wind} < V_{rated} \\ P_{rated} & ;\forall V_{rated} \le V_t^{wind} < V_{c_{out}} \\ 0 & ;\forall V_t^{wind} \ge V_{c_{out}} \end{cases} \tag{40}$$

$$0 \le P_{i,c,t}^{WT} \le P_{i,c,t}^{WTa} \tag{41}$$

6) *Resiliency index*

In this paper, the resiliency index is defined as percent of supplied loads compared to total load as (42). Also, (43) represents the load variation during the day, where $\lambda_t$ is the hourly load factor in per unit.

$$RI = \left(\frac{P^{L_{total}} - OF_2}{P^{L_{total}}}\right) \times 100 \quad ;\forall c \in c_2 \tag{42}$$

$$P^{L_{total}} = \sum_{t \in \Omega_T} \lambda_t \sum_{i \in \Omega_N} P_i^L \tag{43}$$

## IV. Numerical Study

### A. *Test system and input parameters*

The proposed model is applied to the IEEE 33-bus network. The voltage level of this system is 12.66kV, and total active and reactive load of the network at the peak hour are 3.715MW and 2.3MVAr, respectively. The system contains 32 main lines and 5 tie-lines. Twelve lines are equipped with sectionalizes for the aim of reconfiguration and grid formation after fault occurrence. The data of lines and loads have been given in [21]. In addition, Figs. 1 and 2 illustrate the load profile and wind speed pattern during the day, respectively. In Table 1, the required constant parameters of the network equipment have been presented [22].

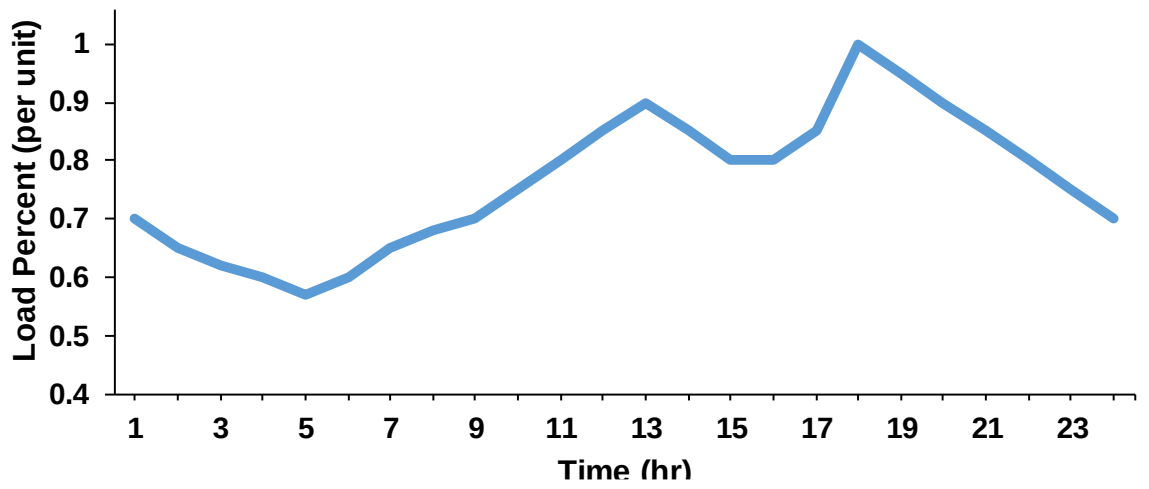

Fig. 1: Daily load profile of the network

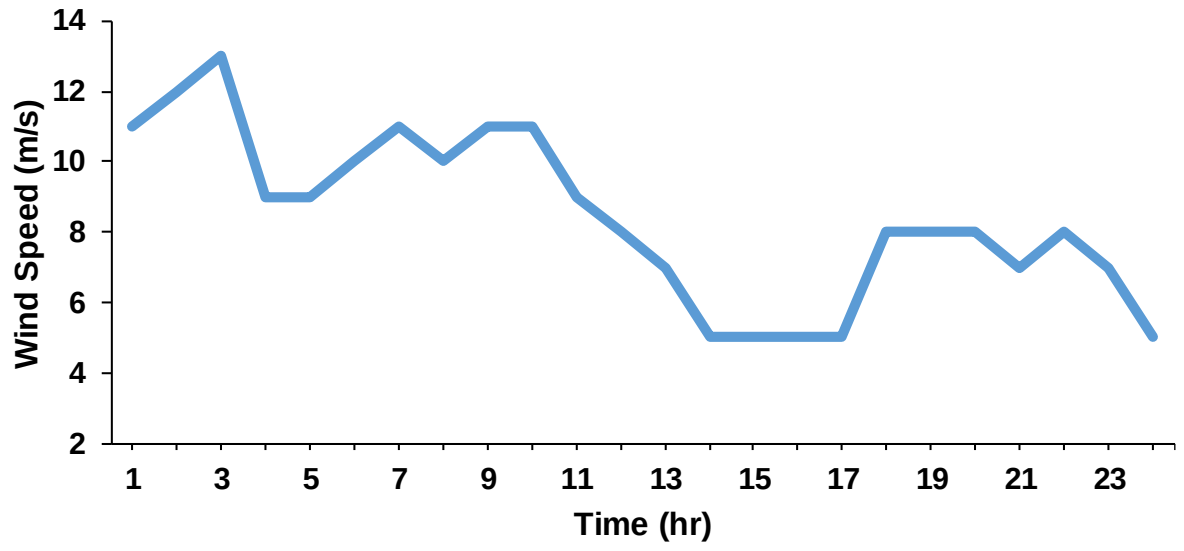

Fig. 2: Wind speed profile during the day

### B. *Optimization process*

The proposed approach in this paper is programmed and implemented in the GAMS software package. As the proposed model is MIQCP, the global optimal solvers can be employed to solve the model. In this paper the "GUROBI" solver has been utilized.

Table 1: Parameters of network equipment

| Element | *Parameter* | *Value* |
|---|---|---|
| WT | $P_{rated}$ | 300 kW |
| | $V_{c_{in}}$ | 3 m/s |
| | $V_{c_{out}}$ | 25 m/s |
| | $V_{rated}$ | 12 m/s |
| ESS | $SOC^{Max}$ | 1.2 MWh |
| | $P_{Max}^{ESS}$ | 0.3 MW |
| | $SOC_{i,c,t_0}$ | 0.4 $SOC^{Max}$ |
| DG | $S_{Max}^{DG}$ | 500 kW |
| | $PF_c \quad ;\forall c \in c_1$ | 0.9 |
| | $PF_c \quad ;\forall c \in c_2$ | 0.8 |
| Network | $I_\ell^{Max}$ | 250 A |
| | Location of WT and ESS units | Buses 5 and 17 |
| | Location of DG units | Buses 21 and 33 |
| | $V_{i,c,t} = 1 \quad ;\forall c \in c_1, i \in \Omega_S$ | |
| | $0.95 \le V_{i,c,t} \le 1.05 \quad ;\forall c \in c_1$ | |
| | $0.9 \le V_{i,c,t} \le 1.1 \quad ;\forall c \in c_2$ | |

### C. *Simulation results*

The proposed approach has been applied to the IEEE 33-bus system in normal and emergency conditions. In this regard, the configuration of network in both operating modes is considered to be radial. It should be noted that after disaster occurrence, the main line connecting the distribution system to its supplying substation will be disconnected and an isolated network is formed. Figs. 3 and 4 depict the schematic view of the constructed network in normal and emergency conditions, respectively. In normal condition (mode $c_1$), one of the tie-lines (line 12-23) has been closed, and one of the main lines (9-10) is switched off to preserve the radial structure. The radiality constraint has also been maintained in emergency condition (mode $c_2$), while the network configuration is different with the normal mode in order to restore the maximum possible loads.

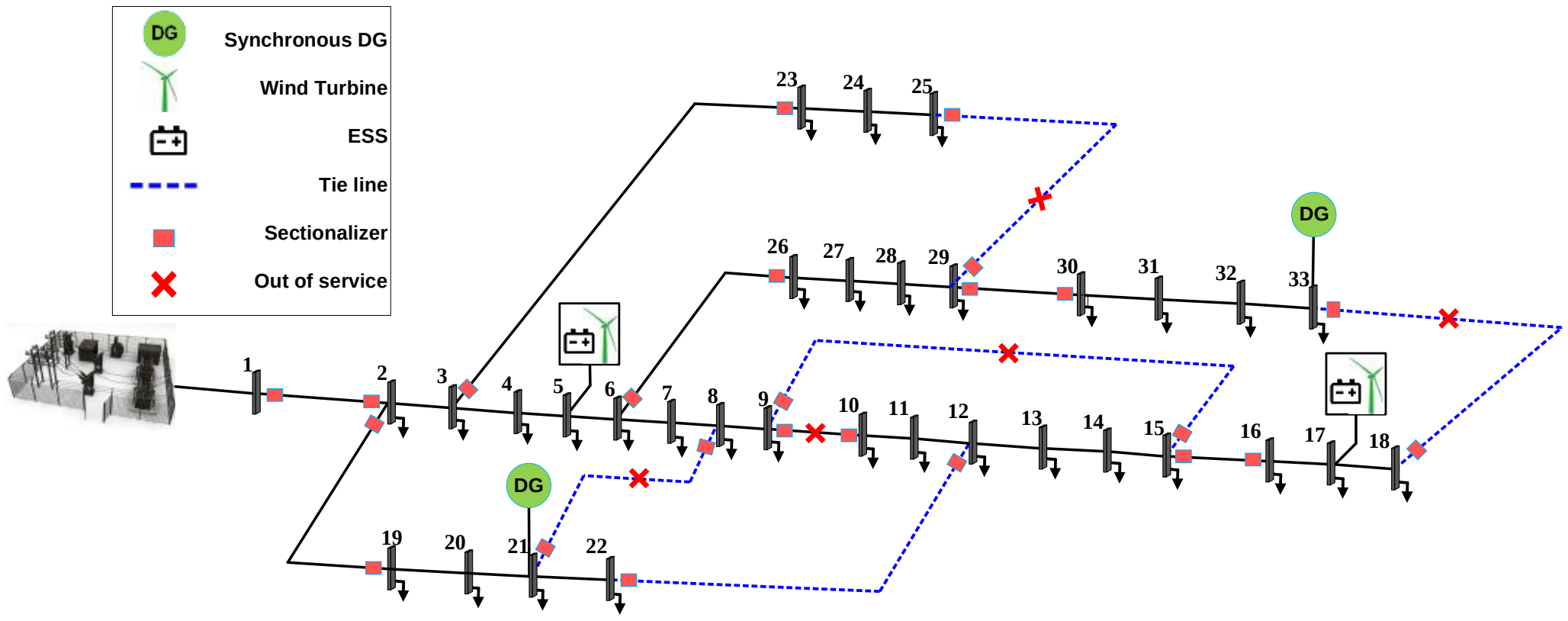

Fig. 3: Schematic of system in case $c_1$

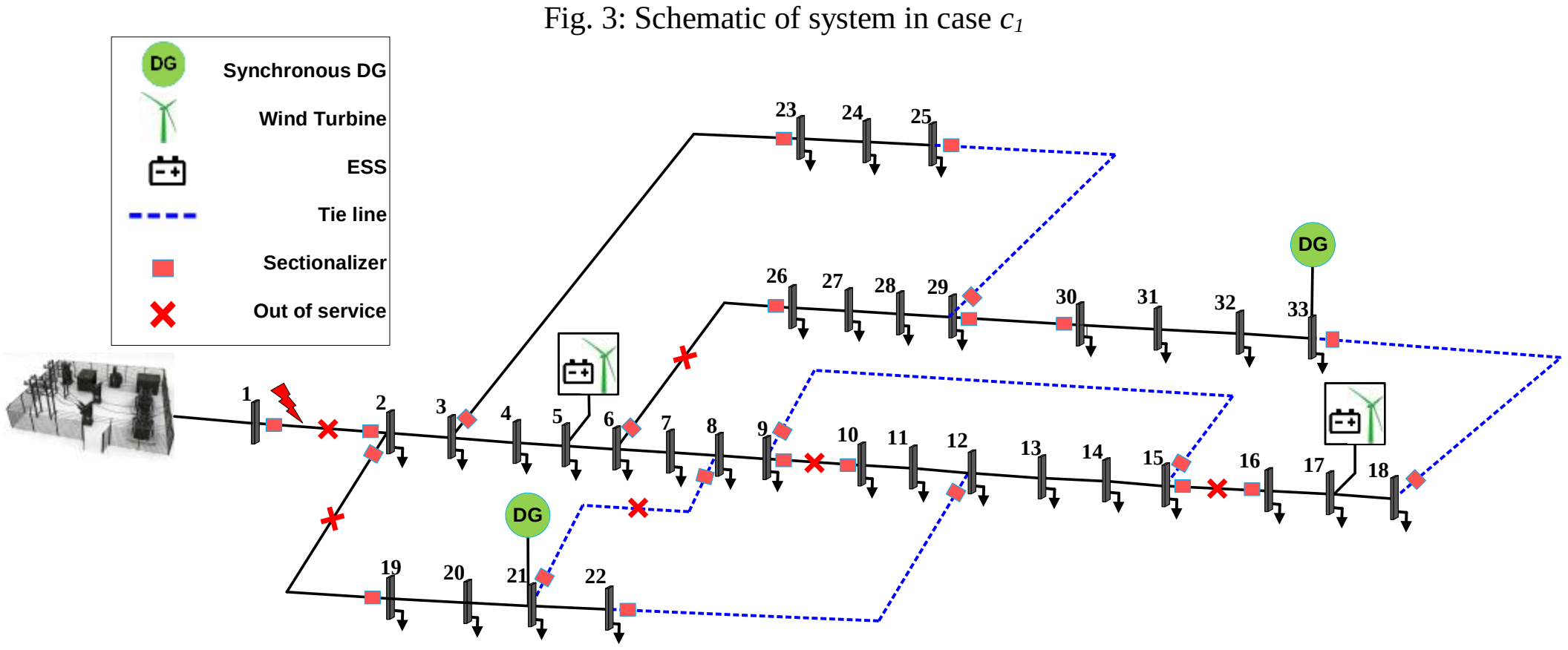

Fig. 4: Schematic of system in case $c_2$

Table 2: Numerical results of the simulation

| $OF_1(MWhr)$ | $OF_2(MWhr)$ | $OF$ | $RI$ | $OF_1^{optimum}(MWhr)$ | $OF_2^{optimum}(MWhr)$ |
|---|---|---|---|---|---|
| 0.905 | 38.879 | 2.000 | 42.875 | 0.905 | 38.860 |

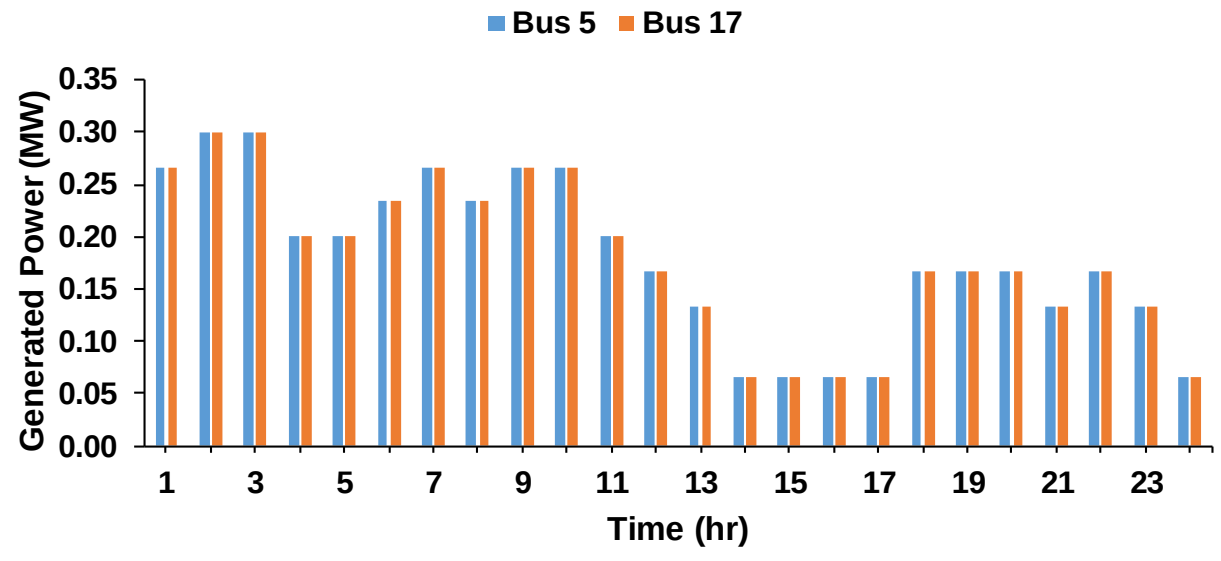

Fig. 5: Active power of wind turbines in mode $c_1$ and $c_2$

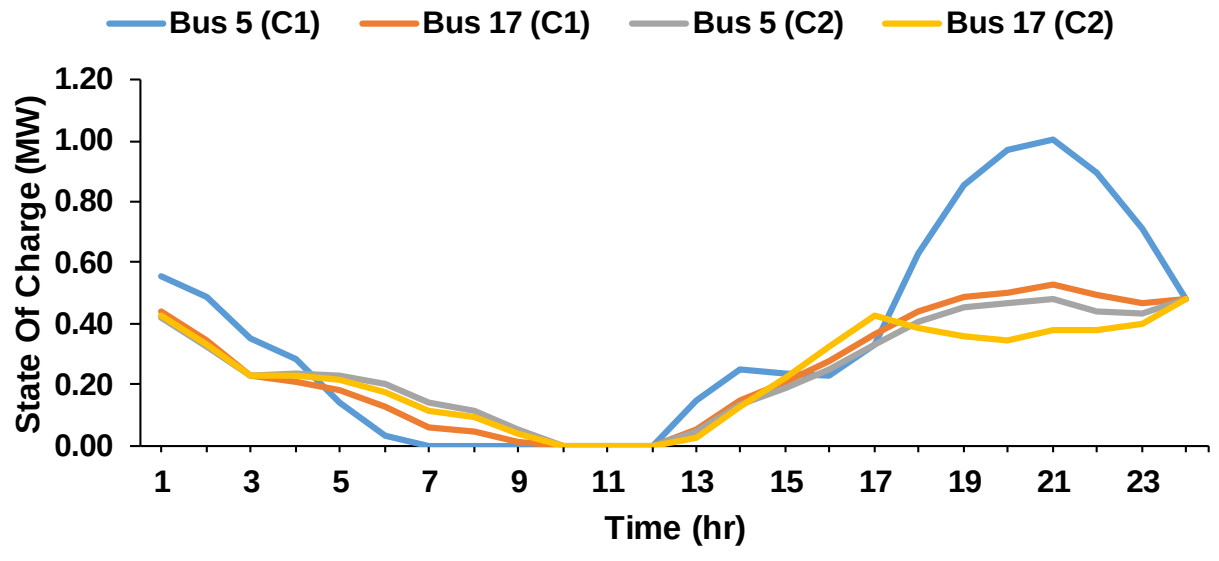

Fig. 6: State-of-charge (energy level) of ESS units in modes $c_1$ and $c_2$

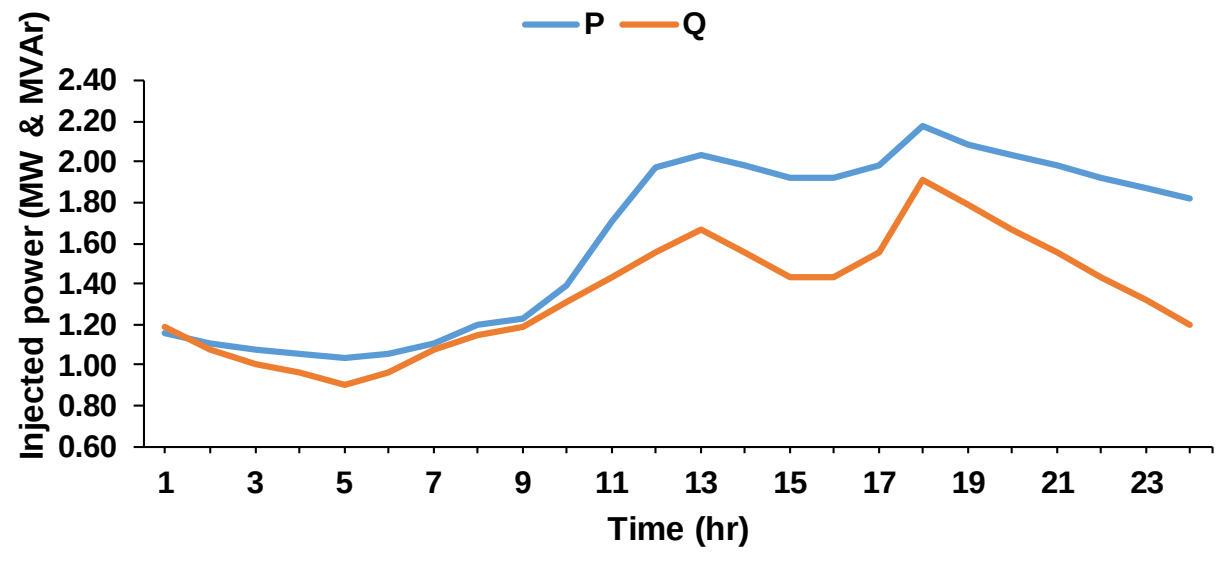

Fig. 7: Active and reactive power injected from upstream network in mode $c_1$

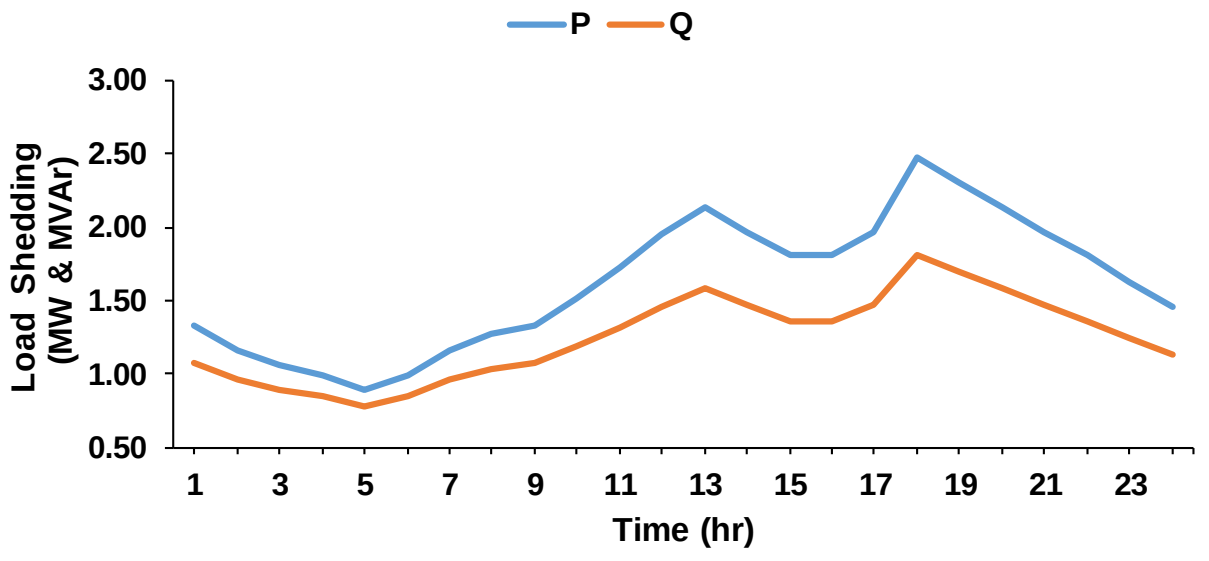

Fig. 8: The total active and reactive load shedding in mode $c_2$

The value of objective function in both normal and isolated condition, and also the resiliency index (*RI*) have been reported in Table 2. Almost 43% of the loads have been restored after the fault occurrence. In Fig. 5, the output power of the wind turbines in different hours is depicted. It is seen that the maximum available power is generated by the wind turbines in both modes. Also, the energy level of ESS units in different hours and modes are shown in Fig. 6. It can be observed that the charging/discharging behavior is different in normal and emergency condtions. Fig. 7 illustrates the injected active and reactive powers from the upward grid into distribution system. As the load demand increases and approaches the peak load hours, the amount of power injection from the upstream network is increased. Also, Fig. 8 shows the total value of load shedding. At the peak hours, due to isolation of the system from the upstream network, the value of load shedding is increased. Finally, Fig. 9 depicts active and reactive powers generated by synchronous DG units for different modes. The generated powers are in a way that the anticipated purpose in each mode i.e. reduction of power loss in normal mode and reduction of load shedding in emergency mode is fulfilled.

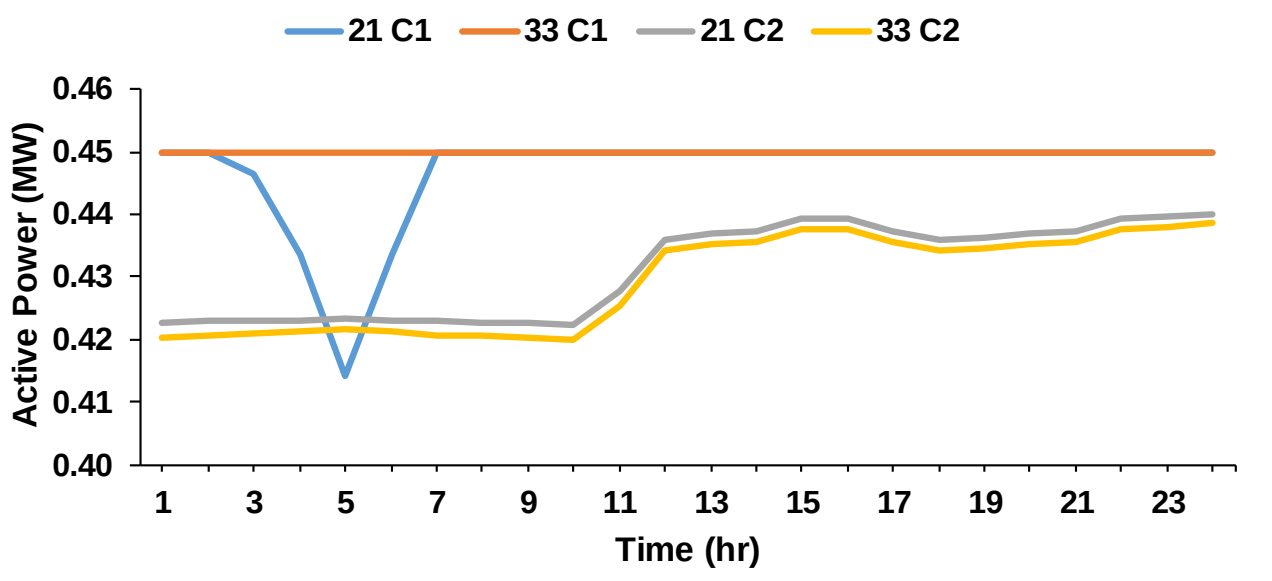

(a)

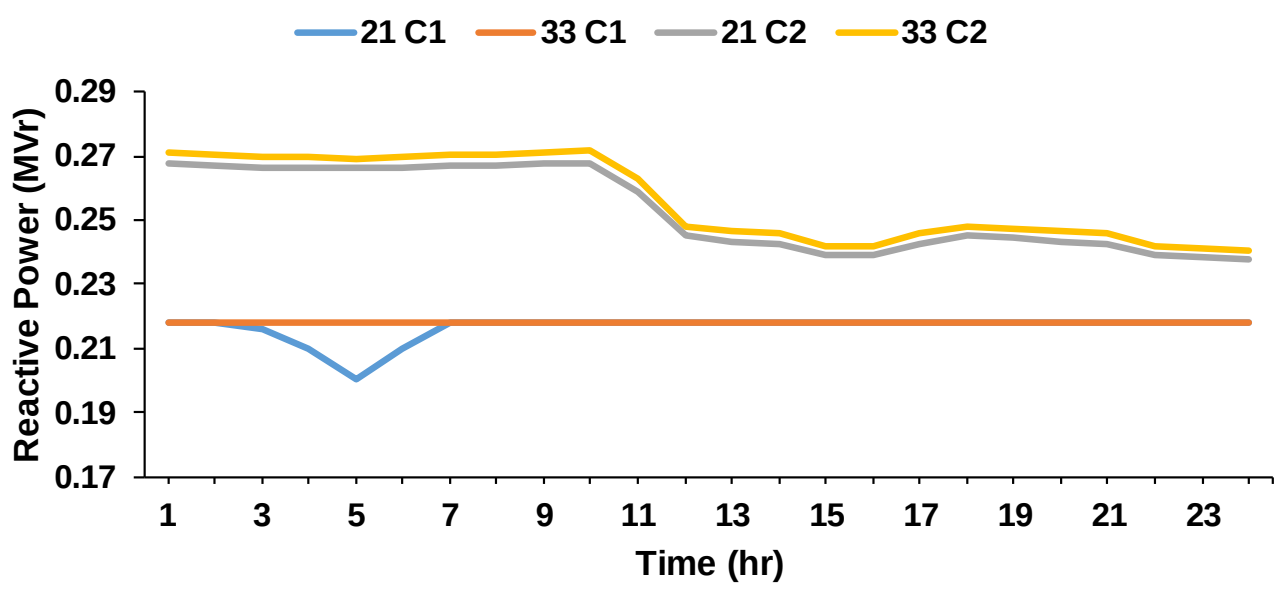

(b)

Fig. 9: Generated active and reactive powers of synchronous DGs in different modes

## V. Conclusion

A new model has been proposed in this paper for optimal operation of ADN considering both normal and emergency conditions. The developed model employs network reconfiguration and optimal grid formation strategies as well as optimal scheduling of DER units to enhance resiliency of the system in critical conditions and minimize the power loss during normal operating condition. The LFB model of AC power flow equations are employed, and all proposed formulations are employed in a linearized form to construct a convex model using MIQCP. The presented approach was tested on the IEEE 33-bus system, and its efficiency has been investigated through different studies. The simulations verify optimal operation of network in both normal and emergency conditions along with satisfying all the problem constraints.